\documentclass[a4paper,10pt]{article}

\usepackage[pdftex]{graphicx}
\pdfoutput=1

\usepackage[T1]{fontenc}
\usepackage[utf8]{inputenc}

\usepackage{fullpage}
\usepackage{natbib}
\usepackage{setspace}
\usepackage[utf8]{inputenc}
\usepackage{amscd,amsmath,latexsym,amssymb,amsfonts}
\usepackage{mathrsfs} 

\usepackage{tikz}
\usetikzlibrary{calc}
\usetikzlibrary{fit}
\usepackage{caption}
\usepackage[list=true]{subcaption}

\usepackage{bm}
\usepackage{amsmath}
\usepackage{color}

\newcommand{\y}{\mathbf{y}}
\newcommand{\s}{\mathbf{s}}
\newcommand{\sobs}{\s_{\text{obs}}}
\newcommand{\yobs}{\y_{\text{obs}}}
\newcommand{\bmu}{\bm{\mu}}

\usepackage{float}
\floatstyle{ruled}
\newfloat{algorithm}{tbp}{loa}
\providecommand{\algorithmname}{Algorithm}
\floatname{algorithm}{\protect\algorithmname}

\usepackage[a4paper,left=2cm,top=2cm,right=2cm,bottom=2cm,ignoreheadfoot]{geometry}

\title{Some discussions of D.~Fearnhead and D.~Prangle's Read Paper ``Constructing summary
statistics for approximate Bayesian computation: semi-automatic
approximate Bayesian computation''}
\author{by\\
Christophe Andrieu$^1$,
Simon Barthelm\'e$^2$, 
Nicolas Chopin$^3$,
Julien~Cornebise$^4$,\\ 
Arnaud Doucet$^{5}$,
Mark~Girolami$^4$,
Ioannis~Kosmidis$^4$.
Ajay Jasra$^6$,\\
Anthony Lee$^{7}$,
Jean-Michel Marin$^8$, 
Pierre Pudlo $^{9,10}$, 
Christian P.  Robert$^{3,11,12}$\\
Mohammed Sedki$^8$ 
and Sumeetpal S. Singh$^{13}$\\
at\\
$^1$University of Bristol,
$^2$Berlin University of Technology,
$^3$CREST-ENSAE,\\
$^4$University College London
$^5$University of Oxford,
$^6$National University of Singapore,\\
$^7$University of Warwick,
$^8$Université Montpellier 2,
$^9$CBGP, INRA, Montpellier,\\
$^{10}$Université Paris-Dauphine, 
$^{11}$IUF,
$^{12}$CREST
and
$^{13}$University of Cambridge
}\date{\today}

\begin{document}
\maketitle 

\begin{abstract}
This report is a collection of comments on the Read Paper of \cite{Fearnhead2011}, to appear in the Journal of
the Royal Statistical Society Series B, along with a reply from the authors.
\end{abstract}

\section{A universal latent variable representation (C. Andrieu, A. Doucet and A. Lee)}

Exact simulation to tackle intractability in model based statistical
inference has been exploited in recent years for the purpose of exact
inference \cite{beaumont2003estimation,beskos2006exact,andrieu2009pseudo,andrieu2010particle}
(see \cite{gourieroux1993indirect} for earlier work). ABC is a specialisation
of this idea to the scenario where the likelihood associated to the
problem is intractable, but involves an additional approximation.
The authors are to be thanked for a useful contribution to the latter
aspect. Our remarks to follow are presented in the ABC context but
apply equally to exact inference. A simple fact which seems to have
been overlooked is that sampling exactly $Y\sim f(y|\theta)$ on a
computer most often means that $Y=\phi(\theta,U)$ where $U$ is a
random vector of probability distribution $D(\cdot)$ and $\phi(\cdot,\cdot)$
is a mapping either known analytically or available as a {}``black-box''.
The vector $U$ may be of random dimension, i.e. $D(\cdot)$ may be
defined on an arbitrary union of spaces (e.g. when the exact simulation
involves rejections), and is most often known analytically - we suggest
to take advantage of this latter fact. In the light of the above one
can rewrite the ABC proxy-likelihood
\[
\tilde{p}(y^{*}|\theta)=\int_{\mathsf{Y}}K(y,y^{*})\times p(y|\theta)dy\quad,
\]
in terms of the quantities involved in the exact simulation of $Y$
\[
\tilde{p}(y^{*}|\theta)=\int_{\mathsf{U}}K\left(\phi(\theta,u),y^{*}\right)\times D(u)du\quad.
\]
In a Bayesian context the posterior distribution of interest is therefore
\[
\tilde{p}(\theta|y^{*})\propto\int_{\mathsf{U}}K\left(\phi(\theta,u),y^{*}\right)\times D(u)du\times p(\theta)\quad.
\]
Provided that $D(\cdot)$ is tractable, we are in fact back to the
usual, analytically tractable, {}``latent variable'' scenario and\textcolor{red}{{}
}\textcolor{black}{any }standard simulation method can be used to
sample $\theta,U$. Crucially one is in no way restricted to the usual
approach where $U_{i}\overset{iid}{\sim}D(\cdot)$ to approximate
the proxy-likelihood. In particular, for $\theta$ fixed, one can
introduce useful dependence between $\phi(\theta,U_{1}),\phi(\theta,U_{2}),\ldots$
e.g. using an MCMC of invariant distribution $D(\cdot)$ started at
stationarity \cite{andrieu2009pseudo}. The structure of $\tilde{p}(\theta,u|y^{*})$
may however be highly complex and sophisticated methods may be required.
One possible suggestion is the use of particle MCMC methods \cite{andrieu2010particle}
to improve sampling on the $U-$space, e.g. for a fixed value of $\theta$
estimate the proxy-likelihood $\int_{\mathsf{U}}K\left(y^{*},\phi(\theta,u)\right)\times D(u)du$
unbiasedly using an SMC sampler \cite{del2006sequential} targeting
a sequence of intermediate distributions between $D(u)$ and $K\left(\phi(\theta,u),y^{*}\right)\times D(u)$
proportional to
\[
K_{j}\left(\phi(\theta,u),y^{*}\right)\times D_{j}(u)
\]
for $\{K_{j}(\cdot,\cdot),j=1,\ldots,n-1\}$ and $\{D_{j}(\cdot),j=1,\ldots,n-1\}$
$ $ and plug such an estimate in standard MCMC algorithms. Notice
the flexibility offered by the choice of $\{K_{j}(\cdot,\cdot)\}$
and $\{D_{j}(\cdot)\}$ which can allow one to progressively incorporate
both the dependence structure on $U$ and the constraint imposed by
$K(\cdot,\cdot)$. When $\phi(\cdot,\cdot)$ is known analytically,
under sufficient smoothness conditions one can use an IPA \cite{pflug1996optimization,andrieu2005line}
approach to estimate e.g. gradients with respect to $\theta$
\[
\nabla_{\theta}\int_{\mathsf{U}}K\left(\phi(\theta,u),y^{*}\right)\times D(u)du\quad.
\]
Again such ideas equally apply to genuine latent variable models and
have the potential to lead to efficient exact inference methods in
otherwise apparently {}``intractable'' scenarios.

\section{Summary-free ABC (S. Barthelm\'e, N. Chopin, A. Jasra and S.S. Singh)}

We strongly believe that the main difficulty with ABC-type methods is the choice of summary statistics.
Although introducing summary statistics may be sometimes beneficial \cite{Wood2010}, in most cases this induces a
bias which is challenging to quantify. We thus welcome this important work on automatically choosing summary
statistics. The fact remains that the optimality criterion proposed in the paper is a bit limiting; we want to
approximate a full posterior distribution, not simply the posterior expectation. In addition, the proposed
approach does not offer a way to monitor the bias induced by the optimal set of summary statistics, except by
numerically comparing many alternative summary statistics, which is potentially tedious.

It is perhaps useful to note there now exist ABC methods that do not use summary statistics, at least for
certain classes of models. The EP-ABC algorithm of \cite{simon_nicolas} is a fast approximation scheme for ABC
posteriors based on constraints on the form $\| y_i-y_i^\star \|<\epsilon$. It is typically orders of magnitude
faster than Monte Carlo based ABC algorithm, whilst, in some scenarios, featuring an approximation error that
is  smaller, due to the absence of summary statistics. It is currently limited however to models such that the
$y_i$ may be simulated sequentially using some chain rule decomposition.

For hidden Markov models, ``exact'' ABC inference (i.e.~not relying on either summary statistics or an
approximation scheme)  may be achieved as well, via the HMM-ABC approach of \cite{dean:singh:jasra:peters:2011,
dean2} (see also \cite{mckinley}), which show that an ABC posterior may be re-interpreted as the posterior of
an artificial hidden Markov model, where the observations are corrupted with noise. This interpretation makes
the remark of \cite{wilkinson} even more compelling: without summary statistics, an ABC posterior may be
interpreted as the correct posterior of a model where the \emph{actual data} (as opposed to the summary
statistics) are corrupted with noise. For instance, the Ricker model example, and with some adaptation for the
Lokta-Volterra example of the read paper.

These two approaches already cover many ABC applications, and could be applied directly to three examples of
the read paper: $g$-and-$k$ distributions (EP-ABC), Lokta-Volterra processes (EP-ABC, HMM-ABC with a slight
modification), and the Ricker model (HMM-ABC).  We are currently working on extending this work in other
dependence structures for the observations and we hope that others will also join us in this effort of removing
summary statistics in ABC.

\section{Inference on summary statistics: a mean or an end? (J.~Cor\-ne\-bi\-se and M.~Girolami)}

We congratulate the authors for their excellent article. We would like to suggest
consideration of cases where it
would make sense, from the perspective of statistical inference, to focus
directly on $p(\theta|\s)$, that is base inferences on the
pre-processed, summarized data $\s$, rather than on the raw data
$\yobs$.
Such a practice is standard in fields such as statistical discriminant
analysis, pattern recognition, machine learning and computer vision, where
pre-processing such as feature extraction \citep[see
e.g.][]{Lowe2004}, edge detection, and thresholding are routine, or in
medical signal processing (e.g. MRI), where inference occurs on pre-processed output
of the medical instrument.
\cite{Wood2010} focuses on qualitative descriptors of noisy chaotic dynamic systems
presenting strong dependence on the initial conditions, with applications to
ecological models: the primary
interest for the user of these models are the characteristics of the
trajectory (regularity, pseudo-period, maxima, extinction of the
population, \ldots), not its actual path.

\begin{figure}[htbp]%
\begin{subfigure}[t]{.45\textwidth}%
\begin{center}
\includegraphics[page=1]{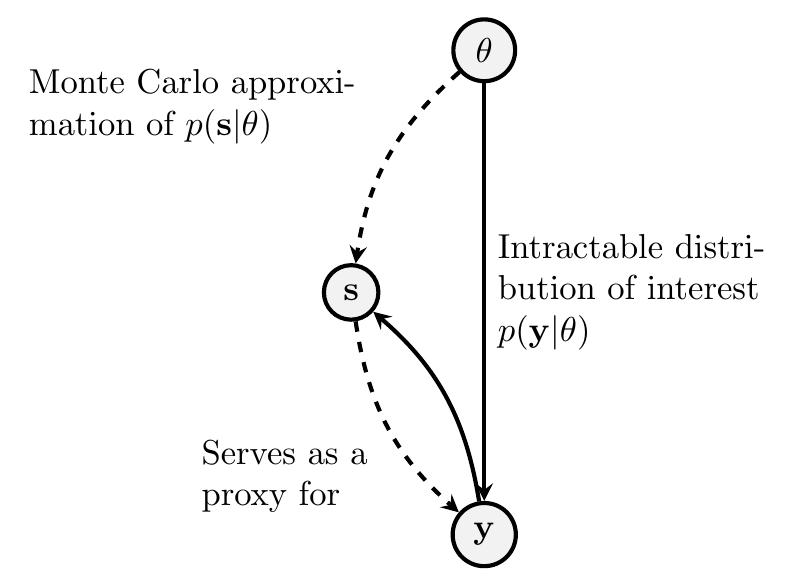}
\end{center}
\caption{Classical use of ABC: inference based on the raw data $\y$, the
summary statistics $\s$ serve to compute a Monte-Carlo estimate of
$p(\s|\theta)$ as a proxy for the intractable likelihood $p(\y|\theta)$.}
\end{subfigure}%
\hfill
\begin{subfigure}[t]{.45\textwidth}
\begin{center}
\includegraphics[page=2]{figure1.pdf}
\end{center}
\caption{Possible complementary use of ABC: inference based on the
summarized data $\s$, the raw data $\y$ serves as an intermediate
simulation step.}
\end{subfigure}
\caption{Graphical representation of the two possible uses of ABC: the
roles of the data $\y$ and of the summary $\s$ are inversed. Plain lines represent
distributions
from which it is easy to sample; Annotated dashed lines represent logical relations.}
\label{fig:dag}
\end{figure}

Statistically speaking, as illustrated in the DAG of Figure~\ref{fig:dag},
this is nothing but shifting the model one layer down the hierarchical
model, permuting the role of $\y$ and $\s$ as auxiliary simulation
variable and variable of interest, with the advantage of removing the
proxy approximation: the summary statistics are not an approximation
anymore, but the actual focus of interest. This is
reminiscent of discriminative–generative modelling (see e.g.
\cite{Xue2010} and \cite{Hopcroft2010}). The
choice of those statistics then becomes either a modelling problem based
on domain specific expertise or, drawing further the comparison with computer
vision, a matter of sparse base construction as recently developed
in compressed sensing \citep{Candes2008}.

The only remaining layer of approximation is that of density estimation by the
kernel $K$. Unfortunately, this kernel density estimation is only
\emph{asymptotically} unbiased, and is biased for \emph{finite} sample
size, the MH ratio in ABC-MCMC \citep[Table~2]{Fearnhead2011} cannot be
cast in the Expected Auxiliary Variable of \cite{Andrieu2007} extending
\cite{andrieu2009pseudo}, not yet available but summarized in \cite{andrieu2010particle}, Section 5.1.

\section{Parametric estimation of the summary statistics likelihood (J.~Cor\-ne\-bi\-se, M.~Girolami and I.~Kosmidis)}  

We would like to draw attention to the work of \cite{Wood2010} which is of
direct relevance to ABC despite it having been largely overlooked in the
ABC literature.
In Section 1.2 of the current paper the authors note that $K\left((S(\y) -
\sobs)/h\right)$ is a Parzen-Rosenblatt density kernel.
As has already been suggested in e.g. \cite{DelMoral2011} one can simulate
$R$ observations $\y_1, \ldots \y_R$ for a given value of $\theta$ and use
the corresponding nonparametric kernel density estimate $\sum_r
K\left((S(\y_r) - \sobs)/h\right)/(R h^d)$ for $p(\sobs|\theta)$.
\cite{Wood2010} suggests the {\em synthetic likelihood} by invoking the assumption of
multivariate normality such that $\sobs \sim \mathcal{N}(\bmu_\theta,
\Sigma_\theta)$. Plug-in estimates of $\bmu_\theta$ and $\Sigma_\theta$ are
obtained by the empirical mean $\hat\bmu^R_\theta$ and covariance $\hat
\Sigma^R_\theta$ using the simulated statistics $S(\y_1), \ldots S(\y_R)$
yielding a parametric density estimate $\mathcal{N}(\sobs;
\hat\bmu^R_\theta, \hat\Sigma^R_\theta)$. This {\em synthetic likelihood}
can then be used in an MCMC setting analagous to MCMC-ABC -- and can
similarly be used in IS-ABC and SMC-ABC settings. The convergence rate  of
the variance of the parametric density estimate is independent of the
dimension of the summary statistics, which is in contrast to the
nonparametric rate which suffers from the curse of dimensionality. This
lower variance could improve mixing of the MCMC algorithm underpinning ABC,
as already demonstrated in the pseudo-marginal approach to MCMC of
\citet{andrieu2009pseudo}.

Of course \cite{Wood2010} does not offer an automatic choice of
the summary statistics: the user selects a (possibly large) set of summary statistics
based on doman knowledge of the problem. This is similar to the way Section 3 offers
to select the ``transformations''
$f(\y)$, which are the first round of summary statistics.
However, the relative weighting of each statistic is
automatically inferred via the corresponding variance estimate. Could such a feature
be of benefit in Semi-automatic ABC?

The assumption of multivariate normality on the distribution of the summary
statistics plays a critical role in Wood's approach. He justifies
it by: \emph{i)} choosing polynomial regression coefficients as summary satistics and,
most interestingly, \emph{ii)} uses a pilot run to improve the normality of the
statistics by 
quantile regression transformations -- a preliminary step conceptually similar to the
pilot ABC run of Section 3.

We conjecture that such transformations could allow for the use of parametric
density estimation within Semi-automatic ABC, possibly benefitting from the
increased convergence rate and making use of the variance of the sampled
statistics. Additionally, we wonder if Theorem 4 could be
modified to study the optimality of such transformed Gaussian statistics.

\section{Automatic tuning of pseudo-marginal MCMC-ABC kernels (A. Lee, C. Andrieu and A. Doucet)}

We congratulate the authors on a structured contribution to the practical
use of ABC methods. We focus here on the conditional joint density
\[
\bar{\pi}_{X,Y|\Theta}(x,y|\theta)=\pi_{Y|\Theta}(x|\theta)\bar{\pi}_{Y|X}(y|x),
\]
which is central to all forms of ABC. Here $x$ and $y$ denote the
simulated and observed data or summary statistics in ABC and $\bar{\pi}_{X|\Theta}=\pi_{Y|\Theta}$.
In the article, $\bar{\pi}_{Y|X}(y|x)=K[(y-x)/h]$ and $\bar{\pi}_{Y|\Theta}(y|\theta)=\int\bar{\pi}_{X,Y|\Theta}(x,y|\theta)\mbox{d}x\neq\pi_{Y|\Theta}(y|\theta)$
leads to the approximation. While neither $\bar{\pi}_{Y|\Theta}(y|\theta)$
nor $\pi_{Y|\Theta}(x|\theta)$ can be evaluated, the ability to sample
according to $\pi_{Y|\Theta}(\cdot|\theta)$ allows for rejection,
importance and MCMC sampling according to $\bar{\pi}_{\Theta,X|Y}(\cdot|y)$.
The calibration of noisy ABC is then immediate. If $\tilde{y}\sim\bar{\pi}_{Y|X}(\cdot|y)$,
then marginally $\tilde{y}\sim\bar{\pi}_{Y|\Theta}(\cdot|\theta^{*})$
since $y\sim\pi_{Y|\Theta}(\cdot|\theta^{*})$ for some $\theta^{*}\in\Theta$.
Inference using $\bar{\pi}$ with $Y=\tilde{y}$ is then consistent
with the data generating process although $\bar{\pi}_{\Theta|Y}(\cdot|\tilde{y})$
may not be closer to $\pi_{\Theta|Y}(\cdot|y)$ than $\bar{\pi}_{\Theta|Y}(\cdot|y)$.

The tractability of $\bar{\pi}_{\Theta,X|Y}$, whose unavailable marginal
$\bar{\pi}_{\Theta|Y}(\cdot|y)$ is of interest puts ABC within the
domain of pseudo-marginal approaches \citep{beaumont2003estimation,andrieu2009pseudo},
and the grouped-independence Metropolis-Hastings (GIMH) algorithm
has been used in \cite{becquet2007new}. We present two novel MCMC-ABC
algorithms based on recent work \citep{adl}, and for simplicity restrict
ourselves to the case $\bar{\pi}_{Y|X}(y|x)\propto\mathbf{1}_{B_{h}(x)}(y)$,
where $\mathbf{1}_{B_{h}(x)}$ is the indicator function of a metric
ball of radius $h$ around $x$. These algorithms define Markov chains
solely on $\Theta$.

In the GIMH algorithm with $N$ auxiliary variables, the state of
the chain is $(\theta,x_{1:N})$ where $x_{1:N}:=\left(x_{1},\ldots,x_{N}\right)$
and at each iteration we propose new values $(\theta',z_{1:N})$ via
$\theta'\sim g(\cdot|\theta)$ and $z_{1:N}\sim\pi_{Y|\Theta}^{\otimes N}(\cdot|\theta')$.
Algorithm \ref{alg:rgimh} presents an alternative to GIMH with the
crucial difference in step $3$, where GIMH would use the previously
simulated values of $x_{1:N}$ instead of sampling $N-1$ new ones.
This algorithm can have superior performance to the GIMH algorithm
in some cases where the latter gets `stuck'. Algorithm \ref{alg:oneh}
involves a random number of simulations instead of fixed $N$, adapting
the computation in each iteration to the simulation problem at hand.
Data is simulated using both $\theta$ and $\theta'$ until a `hit'
occurs. It can be verified that the invariant distribution of $\theta$
is $\bar{\pi}_{\Theta|Y}(\cdot|y)$ for both algorithms. The probability
of accepting the move $\theta\rightarrow\theta'$ after step 1 in
Algorithm \ref{alg:rgimh}, as $N\rightarrow\infty$, approaches 
\[
\min\left\{ 1,\frac{\bar{\pi}_{\Theta|Y}(\theta'|y)g(\theta|\theta')}{\bar{\pi}_{\Theta|Y}(\theta|y)g(\theta'|\theta)}\right\} .
\]
For Algorithm \ref{alg:oneh}, this probability is exactly
\[
\min\left\{ 1,\frac{\pi(\theta')g(\theta|\theta')}{\pi(\theta)g(\theta'|\theta)}\right\} \times\frac{\bar{\pi}_{Y|\Theta}(y|\theta')}{\bar{\pi}_{Y|\Theta}(y|\theta)+\bar{\pi}_{Y|\Theta}(y|\theta')-\bar{\pi}_{Y|\Theta}(y|\theta)\bar{\pi}_{Y|\Theta}(y|\theta')}.
\]
Regarding the {}``automatic'' implementation of ABC, Algorithm \ref{alg:rgimh}
could automate the use of $N$ processors on a parallel computer or
Algorithm \ref{alg:oneh} could be used to automatically adapt computational
effort to the target of interest.

\begin{algorithm}
\caption{Rejuvenating GIMH-ABC}

\label{alg:rgimh}

At time $t$, with $\theta_{t}=\theta$:
\begin{enumerate}
\item Sample $\theta'\sim g(\cdot|\theta)$.
\item Sample $z_{1:N}\sim\pi_{Y|\Theta}^{\otimes N}(\cdot|\theta')$.
\item Sample $x_{1:N-1}\sim\pi_{Y|\Theta}^{\otimes(N-1)}(\cdot|\theta)$.
\item With probability 
\[
\min\left\{ 1,\frac{\pi(\theta')g(\theta|\theta')}{\pi(\theta)g(\theta'|\theta)}\times\frac{\sum_{i=1}^{N}\mathbf{1}_{B_{h}(z_{i})}(y)}{1+\sum_{i=1}^{N-1}\mathbf{1}_{B_{h}(x_{i})}(y)}\right\} 
\]
set $\theta_{t+1}=\theta'$. Otherwise set $\theta_{t+1}=\theta$.\end{enumerate}
\end{algorithm}

\begin{algorithm}
\caption{1-hit MCMC-ABC}

\label{alg:oneh}

At time $t$, with $\theta_{t}=\theta$:
\begin{enumerate}
\item Sample $\theta'\sim g(\cdot|\theta)$.
\item With probability $1-\min\left\{ 1,\frac{\pi(\theta')g(\theta|\theta')}{\pi(\theta)g(\theta'|\theta)}\right\} $,
set $\theta_{t+1}=\theta$ and go to time $t+1$.
\item Sample $z_{i}\sim\pi_{Y|\Theta}(\cdot|\theta')$ and $x_{i}\sim\pi_{Y|\Theta}(\cdot|\theta)$
for $i=1,\ldots$ until $y\in B_{h}(z_{i})$ and/or $y\in B_{h}(x_{i})$.
\item If $y\in B_{h}(z_{i})$ set $\theta_{t+1}=\theta'$ and go to time
$t+1$.
\item If $y\in B_{h}(x_{i})$ set $\theta_{t+1}=\theta$ and go to time
$t+1$.\end{enumerate}
\end{algorithm}

\section{A new perspective on ABC (J.-M. Marin and C.P. Robert)}

In this discussion paper, Fearnhead and Prangle do not follow the usual perspective of looking at ABC as a
converging (both in $N$ and $h$) approximation to the true posterior density
\citep{marin:pudlo:robert:ryder:2011}. Instead, they consider a
randomised (or noisy) version of the summary statistics 
$$ 
s_\text{obs} = S(y_{obs}) + h x\,,\quad x\sim K(x)
$$ 
and they derive a calibrated version of ABC, i.e.~an algorithm that gives ``proper" predictions, but only for
the (pseudo-)posterior based upon this randomised version of the summary statistics. This randomisation however
conflicts with the Bayesian paradigm in that it seems to require adding pure noise to (and removing information
from) the observation to conduct inference. Furthermore, Theorem 2 is valid for any value of $h$. We thus
wonder at the overall statistical meaning of {\em calibration}, since even the prior distribution
(corresponding to $h=+\infty$) is calibrated, while the most informative (or least randomised) case (ABC) is
not necessarily calibrated.  Nonetheless, the interesting aspect of this switch in perspective is that the
kernel $K$ used in the acceptance probability, with bandwidth $h$,
$$ 
K((s-s_\text{obs})/h)\,,
$$
need not behave like an estimate of the true sampling density since it appears in the (randomised) pseudo-model. 

As clearly stated in the paper,  the ABC approximation is a kernel  convolution approximation.  This type of
approximation has been studied in the approximation  theory litterature.  Typically, \cite{light:1993} introduces a
technique for generating an  approximation to a given continuous function using convolution kernels. Also, in
\cite{levesley:etal:1996}, it is constructed a class of continuous integrable functions to serve as kernels
associated with convolution operators that produce approximations to arbitrary continuous  functions. It could
be eventually promising to adapt some of the techniques introduce in these papers.

Overall, we remain somehow skeptical about the ``optimality" resulting from this choice of summary statistics
as (a) practice---at least in population genetics \citep{cornuet:santos:beaumont:etal:2008}---shows that proper
approximation to genuine posterior distributions stems from using a  number of summary statistics that is
(much) larger than the dimension of the parameter; (b) the validity of the approximation to the optimal summary
statistics used as the actual summary statistics ultimately depends on the quality of the pilot run and hence
on the choice of the summary statistics therein; this approximation is furthermore susceptible to deteriorate
as the size of the pilot summary statistics grows; (c) important inferencial issues like model choice are not
covered by this approach and recents results of ours \citep{marin:pillai:robert:rousseau:2011} show that
estimating statistics are likely to bring inconsistent solutions in this context; those results imply
furthermore than a naïve duplication of Theorem 3, namely based on the Bayes factor as a candidate summary
statistic, would be most likely to fail.

In conclusion, we congratulate the authors for their original approach to this major issue in ABC design and,
more generaly, for bringing this novel and exciting inferential method to the attention of the readership.

\section{On the consistency of noisy ABC (C.P. Robert)}

A discussion paper on the fast-growing technique of ABC techniques is quite timely, especially when it
addresses the important issue of summary statistics used by such methods. I thus congratulate the authors on
their endeavour. 

While ABC has been gradually been analysed from a (mainstream) statistical perspective, this is one of the very
first papers performing a decision-theoretic analysis of the factors influencing the performances of the method
(along with, e.g., \citealp{dean:singh:jasra:peters:2011}). Indeed, a very interesting input of the authors is
that ABC is considered there from a purely inferential viewpoint and calibrated for estimation purposes.  The
most important result therein is in my opinion the consistency result in Theorem 2, which shows that noisy
ABC is a coherent estimation method when the number of observations grows to infinity. I however dispute the
generality of the result, as explained below.

In Fearnhead's and Prangle's setting, the Monte Carlo error that is inherent to ABC is taken into 
account through the average acceptance probability, which collapses to zero when $h$ goes to zero, 
meaning that $h=0$ is a suboptimal choice. This is a strong (and valid) point of the paper because this means that the
``optimal" value of $h$ is not zero, a point repeated later in this discussion. The later decomposition of the
error into 
$$
\text{trace}(A\Sigma) + h^2 \int x^\text{T} A x K(x) \text{d} x + \dfrac{C_0}{Nh^d}
$$
is very similar to error decompositions found in (classical) non-parametric statistics.  In this respect, I do
fail to understand the argument of the authors that Lemma 1 implies that a summary statistics with larger
dimension also has larger Monte Carlo error: Given that  $\pi(s_\text{obs})$ also depends on $h$, the
appearance of $h^d$ in eqn.~(6) is not enough of an argument. There actually is a larger issue I also have
against several recent papers on the topic, where the bandwidth $h$ or the tolerance $\epsilon$ is treated as a
given or an absolute number while it should be calibrated in terms of a collection of statistical and computational
factors, the number $d$ of summary statistics being one of them.

When the authors consider the errors made in using ABC, balancing the Monte
Carlo error due to simulation with the ABC error due to approximation (and non-zero tolerance), they fail to
account for ``the third man" in the picture, namely the error made in replacing the (exact) posterior inference
based on $\mathbf{y}_\text{obs}$ with the (exact) posterior inference based on $\mathbf{s}_\text{obs}$,
i.e.~for the loss of information due to the use of the summary statistics at the centre of the Read Paper.
(As shown in \citealp{robert:cornuet:marin:pillai:2011}, this loss may be quite extreme as to the resulting
inference to become inconsistent.) While the remarkable (and novel) result in the proof of Theorem 3 that
$$
\mathbb{E}\{\theta|\mathbb{E}[\theta|\mathbf{y}_\text{obs}]\} =
\mathbb{E}[\theta|\mathbf{y}_\text{obs}]
$$
shows that $\mathbf{s}_\text{obs}=\mathbb{E}[\theta|\mathbf{y}_\text{obs}]$ does not loose any (first-order)
information when compared with $\mathbf{y}_\text{obs}$, hence is ``almost" sufficient in that weak sense, Theorem 3
only considers a specific estimation aspect, rather than full Bayesian inference, and is furthermore
parameterisation dependent. In addition, the second part
of the theorem should be formulated in terms of the above identity, as ABC plays no role when $h=0$.

If I concentrate more specifically on the mathematical aspects of the paper, a point of the utmost importance
is that Theorem 2 can only hold at best when $\theta$ is identifiable for the distribution
$\mathbf{s}_\text{obs}$. Otherwise, some other values of $\theta$ satisfy
$p(\theta|\mathbf{s}_\text{obs})=p(\theta_0|\mathbf{s}_\text{obs})$. Considering the specific case of an
ancilary statistic $\mathbf{s}_\text{obs}$ clearly shows the result cannot hold in full generality.  Therefore,
vital assumptions are clearly missing to achieve a rigorous formulation of this theorem. The call to
\citealp{bernardo:smith:1994} is thus not really relevant in this setting as the convergence results therein
require conditions on the likelihood that are not necessarily verified by the distribution of
$\mathbf{s}_\text{obs}$. We are thus left with the open question of the asymptotic validation of the noisy ABC
estimator---ABC being envisioned as an inference method {\em per se}---when the summary variables are not
sufficient. Obtaining necessary and sufficient conditions on those statistics as done in
\cite{marin:pillai:robert:rousseau:2011} for model choice is therefore paramount, the current paper obviously
containing essential features to achieve this goal.

In conclusion, I find the paper both exciting and bringing both new questions and new perspectives to the
forefront of ABC research. I am thus unreservedly seconding the vote of thanks.

\section{On selecting summary statistics by post-processing (M. Sedki and P. Pudlo)}

\noindent We congratulate the authors for their interesting and
stimulating paper on ABC. Our attention was drawn to the 
regression building the new statistics in Section 3. Fearnhead
and Prangle point similarities with the post-processing proposed by
\citet{beaumont:zhang:balding:2002}.  
But 
they defend their algorithm on its ability to select an efficient
subset of summary statistics.  The main idea here is certainly to
bypass the curse of dimensionality. E.g., in population genetics, a
large number of populations commonly induce more than a hundred summary
statistics with the DIYABC software of
\citet{cornuet:santos:beaumont:etal:2008}.

Apart from \citet{blum:2010}, the widely used post-processing of
\citet{beaumont:zhang:balding:2002} has been little studied
theoretically, although it significantly improves the accuracy of the
ABC approximation.  Actually, \citet{beaumont:zhang:balding:2002}
replace the $\theta$'s kept in the rejection algorithm with the
residuals of a regression learning $\theta$ on the summary
statistics. In the model choice settings \citep[see,
e.g.,][]{robert:cornuet:marin:pillai:2011}, this post-processing uses a logistic regression
predicting the model index, see \citet{beaumont:2008}. In both cases,
it attempts to correct the discrepancy between the observed dataset
and the simulated ones accepted by the ABC algorithm.  We were
intrigued by what would happen when postponing the variable selection
criterion proposed in this paper until this post-processing.

\begin{figure}[htb]
  \centering
  \includegraphics[width=.46\linewidth, height=.46\linewidth]{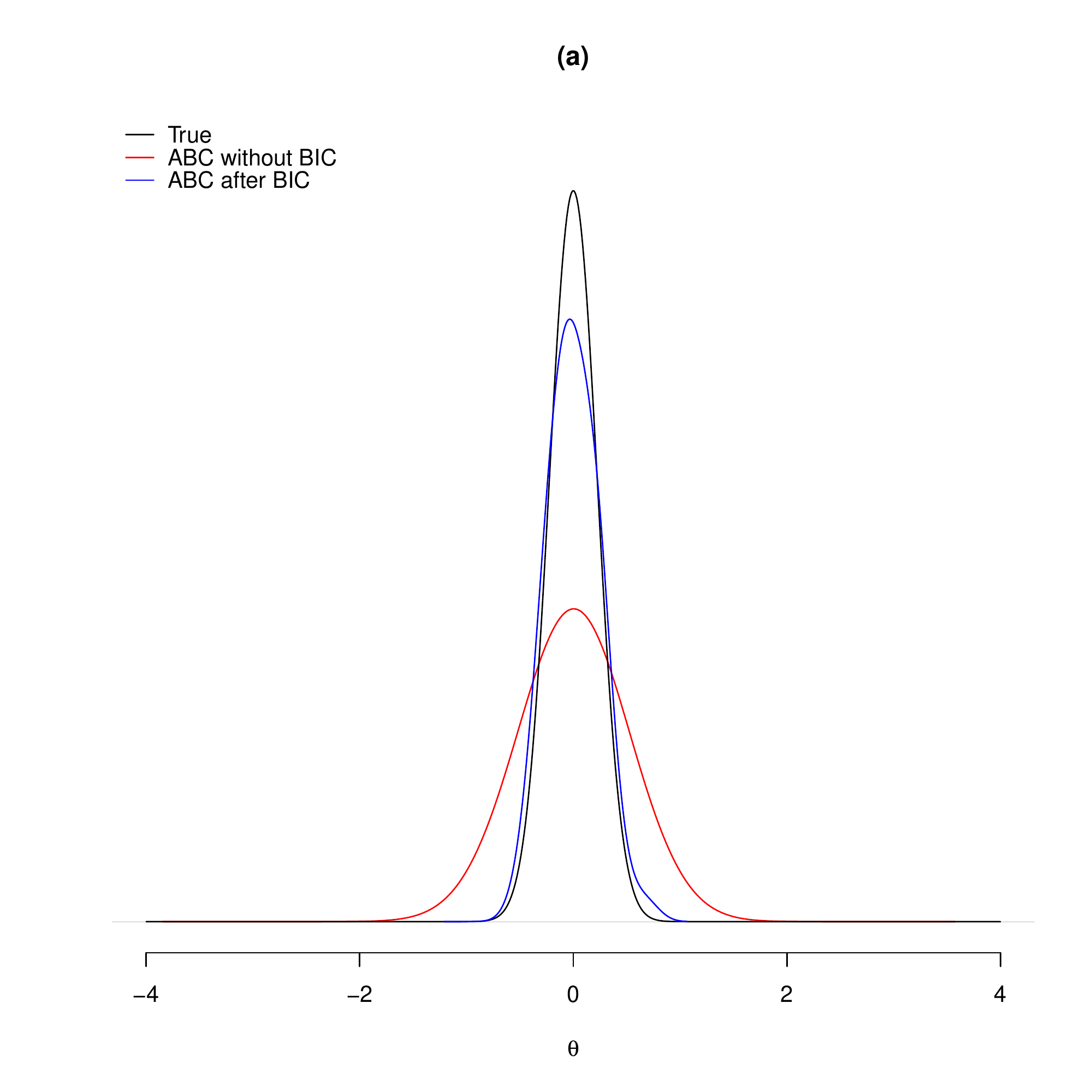}
  \includegraphics[width=.46\linewidth, height=.46\linewidth]{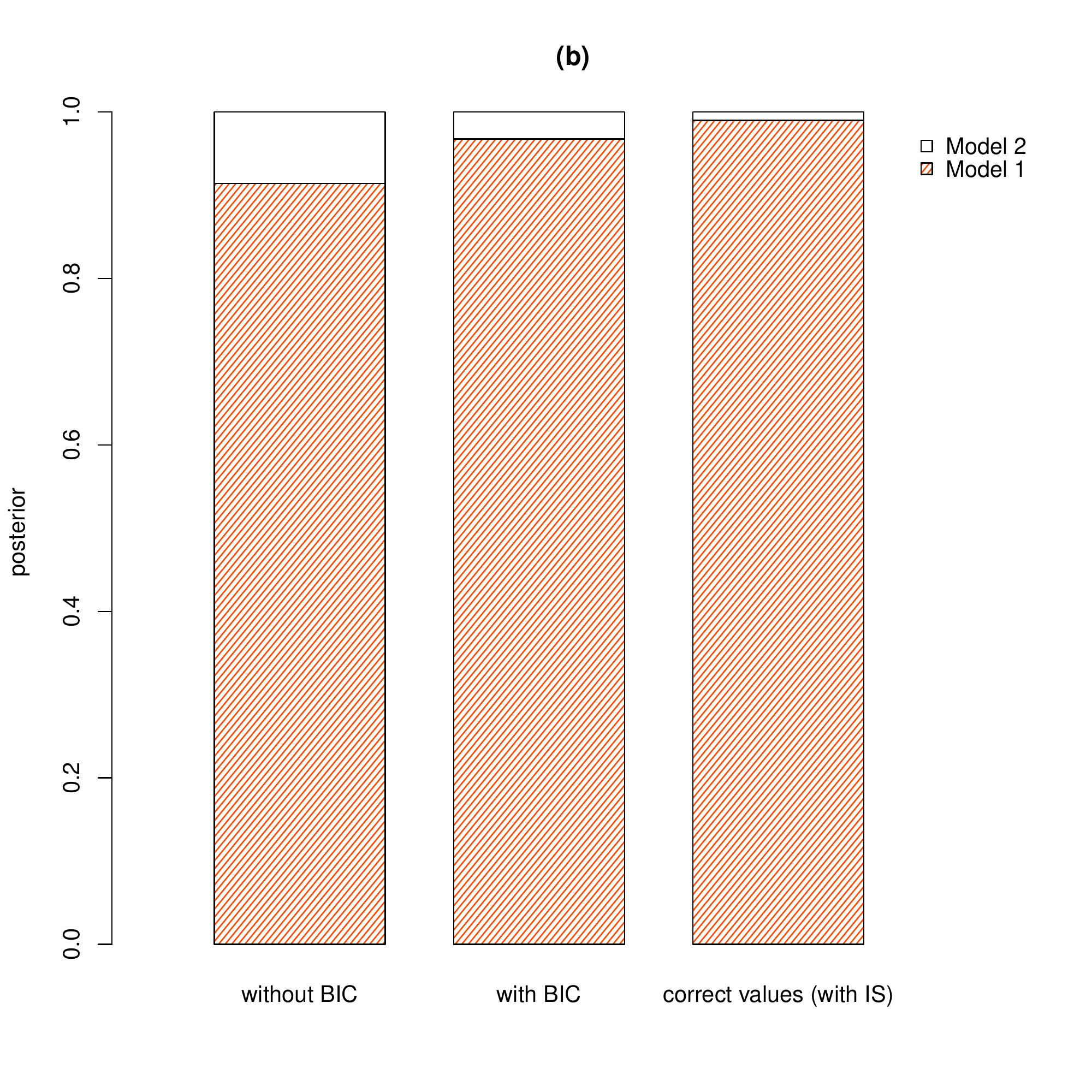}
  \caption{\small %
    \textbf{(a)} Posterior density estimates in the first example. Prior over
    $\theta$ is $\text{Unif}(-5,5)$, while $X$ is a Gaussian vector of
    dimension $20$, with independant components, $X_i|\theta \sim
    \mathcal{N}(\theta, 1)$. Summary statistics are
    $S_1=\text{mean}(X_{1:20})$, $S_2=\text{median}(X_{1:20})$,
    $S_3\sim \text{Unif}(-5, 5)$ and $S_4 \sim \mathcal
    N(0,1)$. Applying BIC here impoves the posterior density estimates
    by removing $S_3$ and $S_4$. %
    \textbf{(b)} The model choice problem which is described in
    \citet{robert:cornuet:marin:pillai:2011} might be summed up the following way:
    considering three populations, we have to decide whether population
    3 diverged from population 1 (Model 2) or 2 (Model 1). Among 24
    summary statistics, BIC selects
    the two summary statistics LIK31 and LIK32 (see Tab. S1 of
    \citet{robert:cornuet:marin:pillai:2011})  which estimates
    genetic similarities between population 3 and the two other ones.
  }
  \label{fig:sedki_pudlo}
\end{figure}

Although a more detailed study is needed, we implemented two
experiments: \textbf{(a)} one with a parameter estimation in the
Gaussian family and \textbf{(b)} one with a model choice in the first
population genetics example of \citet{robert:cornuet:marin:pillai:2011}. We ran the
classical ABC algorithm and used a Bayesian information criterion
(BIC) during the local linear regression to select the relevant
statistics. Then, we scanned once again the whole reference table
drawn from the prior to find the nearest particles to the observation,
considering only the subset of statistics selected by BIC. We ended
with a local linear regression on this new set of
particles. 
Numerical results are given in Figure~\ref{fig:sedki_pudlo} and show
that applying BIC during \citet{beaumont:zhang:balding:2002}'s
post-processing is a promising idea.

\end{document}